\documentclass[12pt,dvips]{article}
\usepackage[dvips]{graphicx}

\def\Ca{{\rm Ca}^{2+}}
\def\Na{{\rm Na}^{+}}
\def\Cacon{[{\rm Ca}^{2+}]}
\def\Nacon{[{\rm Na}^{+}]}
\def\Caconin{[{\rm Ca}^{2+}]_{\rm i}}
\def\Caconout{[{\rm Ca}^{2+}]_{\rm o}}
\def\Naconin{[{\rm Na}^{+}]_{\rm i}}
\def\Naconout{[{\rm Na}^{+}]_{\rm o}}
\def\ncxer{$\Na/\Ca$ exchanger}
\def\ncxers{$\Na/\Ca$ exchangers}
\def\ncx{$\Na/\Ca$ exchange}
\def\sm{smooth muscle}
\def\smc{smooth muscle cell}
\def\smcs{smooth muscle cells}
\def\sr{sarcoplasmic reticulum}

\def\js{junctional space}
\def\jss{junctional spaces}
\def\jn{junction}
\def\jns{junctions}

\def\oneun#1{{\rm #1}}

\def\nm{{\rm nm}}
\def\mum{\mu{\rm m}}
\def\mus{\mu{\rm s}}

\def\m{{\rm m}}

\def\mumol{\mu{\rm mol}}

\def\muM{\mu{\rm M}}

\def\beq{\begin{equation}}
\def\endeq{\end{equation}}
\def\bt{\begin{table}}
\def\endt{\end{table}}
\def\bfig{\begin{figure}}
\def\endfig{\end{figure}}

\long\def\symbolfootnote[#1]#2{\begingroup%
\def\thefootnote{\fnsymbol{footnote}}\footnote[#1]{#2}\endgroup}

\begin{document}
\thispagestyle{empty}
\baselineskip 7.5 mm

\begin{center}
{\bf {\large A quantitative model for linking \ncx\ to SERCA during
refilling of the
    \sr\ to sustain  $\Cacon$ oscillations in vascular smooth muscle}}
\vspace{3mm}
    
\vspace{5 mm}

\noindent Nicola Fameli\hspace{10mm} Cornelis van
Breemen\symbolfootnote[2]{Contributed equally to this
article.}\symbolfootnote[3]{Corresponding author. email: {\tt
breemen@interchange.ubc.ca}}\hspace{10mm} Kuo-Hsing Kuo$^\dag$\\
{\it Department of Anesthesiology, Pharmacology and Therapeutics} \\ 
{\it The University of British Columbia}\\
{\it 2176, Health Sciences Mall, Vancouver, B. C., Canada V6T 1Z3\\
tel 1-604-8225565; fax 1-604-8222281}
\end{center}

\pagebreak

{\centerline{\Large Abstract}}
\vspace{5mm}

We have developed a quantitative model for the creation of 
cytoplasmic $\Ca$ gradients near the inner surface of the plasma
membrane (PM). In particular we simulated the refilling of
the sarcoplasmic reticulum (SR) via PM-SR junctions during
asynchronous $\Caconin$
oscillations in smooth muscle cells of the rabbit {\it inferior vena
cava}. We have combined confocal microscopy data on the $\Caconin$
oscillations, force transduction data from cell contraction studies
and electron microscopic images to build a basis for computational
simulations that model the transport of calcium ions from \ncxers\
(NCX) on
the PM to sarcoplasmic/endoplasmic reticulum $\Ca$
ATPase (SERCA) pumps on the SR as a
three-dimensional random walk through the PM-SR junctional
cytoplasmic spaces.
Electron microscopic ultrastructural images of the smooth muscle
cells were
elaborated with software algorithms to produce a very clear and
dimensionally accurate picture of the PM-SR junctions. From this
study, we
conclude that it is plausible and possible for enough $\Ca$ to pass
through the PM-SR junctions to replete the SR during the regenerative
$\Ca$ release, which underlies agonist induced asynchronous $\Ca$
oscillations in vascular smooth muscle.
\vspace{3mm}

\pagebreak

\section{INTRODUCTION}
There is considerable current interest in the mechanisms whereby the
intracellular messenger $\Ca$ is able to signal different cellular
functions, such as contraction, proliferation, migration and
apoptosis as dictated by variable physiological stimuli or
pathological insult~\cite{casey06}. 
This plasticity of $\Ca$ signaling is
dependent on the coordinated activity of the various $\Ca$ channels,
pumps and exchangers located in the PM,
SR and mitochondria to generate transient or
steady-state cytoplasmic $\Ca$ gradients near $\Ca$-sensitive enzymes
and ion channels~\cite{berridge03}.
The essence of this control system is an
ultrastructural arrangement of the organelles that allows $\Ca$ to
flow from a source, e.g., a channel or \ncxer\ (NCX) to a $\Ca$ sink,
e.g.,
sarcoplasmic/endoplasmic reticulum $\Ca$
ATPase (SERCA) across a cytoplasmic microdomain with restricted
diffusional
characteristics, which prevent equilibration with the bulk of the
cytoplasmic $\Ca$~\cite{casey77, damon04, rizzuto06}.
In this communication, we present a specific quantitative model for 
this type of linked $\Ca$ transport, which serves to refill
the SR during agonist-induced waves of SR $\Ca$ release in vascular
smooth muscle. However, it is implied that similar mechanisms may
underlie the generation of different sub-cellular $\Ca$ gradients
involved in site and function specific $\Ca$ signaling in general.

Iino and coworkers~\cite{iino94} discovered that activation of large
arteries involved the
generation and maintenance of repetitive asynchronous $\Ca$ waves
along the length of smooth muscle cells. In a subsequent series of
papers, we have
investigated their mechanisms in the {\it inferior vena cava} (IVC)
of the rabbit and
elucidated the following serial events (reviewed in~\cite{lee02-2}):
1) activation of IP$_3$ receptor channels as the first event in
the signaling sequence of phenylephrine-mediated $\Caconin$
oscillations leading to \sm\ contraction;
2) activation of non-selective cation channels (NSCC; e.g., receptor-
and
store-operated channels, ROCs/SOCs), causing mainly influx of $\Na$
into the junctional cytosol to
facilitate operation of NCX in the $\Ca$ influx
mode;
3) $\Ca$ influx through reverse NCX;
4) opening of L-type voltage gated $\Ca$ channels (VGCCs);
5) SERCA
pumping $\Ca$ into the lumen of the junctional SR (jSR) to replenish
$\Ca$ lost during the initial release.

Our previous findings revealed that
the bulk of $\Ca$ reloading of the SR during these repetitive $\Ca$
waves is mediated by the reversal of
the NCX linked to $\Ca$ uptake into the SR by SERCA. We postulated
that this
process required junctional complexes of closely apposed PM
and SR membranes separated by a narrow space characterized by
restricted diffusion of calcium ions. 
In support of the requirement of PM-SR junctions for the
maintenance of $\Ca$ oscillations was the disappearance of the latter
upon experimental separation of the two membranes~\cite{lee05}.
However, general
acceptance of the functional linkage between NCX and SERCA requires,
besides experimental evidence, a quantitative demonstration that
efficient, linked $\Ca$ transport between reverse NCX in the PM and
SERCA in the apposing SR is indeed plausible. We therefore
carefully re-examined the ultrastructural characteristics of the
PM-SR junctions in the rabbit IVC by electron microscopy,
estimated from the literature what the magnitude of the junctional
fluxes would have to be in order to fulfill their function of
stimulating contraction and
applied these geometric and dynamic parameters to a mathematical
model based on random walk simulations. The close correlation between the
computed fluxes and experimental data led to the conclusion that PM-SR
junctional $\Ca$ transport could function efficiently to refill the
SR to maintain excitatory $\Ca$ waves in vascular smooth muscle.

\section{METHODS}\label{methods}
\subsection{Tissue preparation}
All the experiments and procedures were carried out in accordance
with the guidelines of the University of British Columbia Animal Care
Committee (protocol number: A990290). Detailed methods have been
previously described for this preparation in
references~\cite{lee02-2} and~\cite{kuo03}. Male New Zealand White
rabbits (1.5--2.5~kg, obtained from Animal Care, University of
British Columbia) were sacrificed by CO$_2$ asphyxiation and then
exsanguinated. The IVC was removed, cleaned of surrounding connective
tissue and then inverted. The endothelium was removed by gently
wiping it with filter paper and the inverted vessel was then
dissected into multiple ring segments that were 4~mm in width.

\subsection{Confocal $\Caconin$ measurements}
The details of the confocal $\Ca$ imaging method have been described
in previous work~\cite{confocdetails}. Briefly, the inverted IVC
rings were loaded with
fluo-4 AM ($5\,\muM$, with $5\,\muM$
pluronic F-127) for 90 min at 25~$^{\circ}$C and then left to
equilibrate for 30 minutes in normal
PSS. The changes in $\Caconin$ were measured using an inverted Leica
TCS SP2 AOBS, laser scanning confocal microscope with an air 10X
(numerical aperture 0.3) lens. The tissue was illuminated using the
488~nm line of an Ar-Kr laser and a high-gain photomultiplier tube
collected the emission at wavelengths between 505~nm and 550~nm. The
acquisition rate was 3~frame/second. The measured changes in Fluo-4
fluorescence level are proportional to the relative changes in
$\Caconin$. All parameters (laser intensity, gain, etc.) were
maintained constant during the experiment. The confocal images were
further analyzed by using the Image J open source software
package~\cite{imagej}.

\subsection{Electron microscopy study}\label{TEMmethods}
Details of the electron microscopy study have been described
previously~\cite{temdetails}. Briefly, for the conventional
microscopy study, the primary fixative solution contained 1.5\%
glutaraldehyde, 1.5\% paraformaldehyde and 1\% tannic acid in 0.1~M
sodium cacodylate buffer that was pre-warmed to the same temperature
as the experimental buffer solution (37~$^{\circ}$C). The rings of
rabbit IVC were fixed at 37~$^{\circ}$C for 10 minutes, then
dissected into small blocks, approximately
1~mm$\times$0.5~mm$\times$0.2~mm in size and put in the same fixative
for 2 h at 4~$^{\circ}$C on a shaker. The blocks were then washed
three times in 0.1~M sodium cacodylate (30 min). In the process of
secondary fixation, the blocks were post-fixed with 1\% OsO4 in 0.1~M
sodium cacodylate buffer for 1 h followed by three washes with
distilled water (30 min). The blocks were then further treated with
1\% uranyl acetate for 1 h ({\it en bloc} staining) followed by three
washes with distilled water. Increasing concentrations of ethanol
(25, 50, 70, 80, 90 and 95\%) were used (10 min each) in the process
of
gradient dehydration. 100\% ethanol and propylene oxide were used
(three 10 min washes each) for the final process of dehydration. The
blocks were infiltrated in the resin (TAAB 812)
and then embedded in molds and polymerized in an oven at
60~$^{\circ}$C for 8 h. The embedded blocks were serial-sectioned on
a microtome using a diamond knife at the thickness of 80~nm. The
serial sections were then stained with 1\% uranyl acetate and
Reynolds lead citrate for 4 and 3 min, respectively. Images were
obtained with a Phillips 300 electron microscope at 80~kV. The images
of serial sections were further processed using Adobe Photoshop
software. The PM and SR
were delineated and pseudo-colored by red channel and green channel,
respectively.  The serial images are further rendered and processed
using the Volume J plugin in the Image J software~\cite{imagej}.

For the immuno-electron microscopy study, in order to preserve
antigenicity, the fixative was replaced with 2.5\% paraformaldehyde
in 0.1 M phosphate buffer saline (PBS) and post-fixation with OsO$_4$
was avoided.  After fixation for 2 h at 4$^{\circ}$C, the blocks were
washed with 0.1 M PBS and followed with three washes with distilled
water. The blocks were dehydrated with increasing concentrations of
ethanol (25, 50, 60 and 70\%) at 4$^{\circ}$C. The blocks were then
infiltrated and embedded in hydrophilic resin (LR white, medium
hardness). The embedded blocks were sectioned and collected for
immuno-labeling. The sections were incubated in 10\% goat serum at
37$^{\circ}$C for 1 h and then incubated with antibodies against NCX
overnight at 4$^{\circ}$C.  After three washes of PBS, sections were
labeled with secondary antibody conjugated with 10 nm colloidal gold
for 2 h at 37$^{\circ}$C.  The labeled sections were stained with 1\%
uranyl acetate and Reynolds lead citrate for 20 and 10 min,
respectively. Images were obtained with a Phillips 300 electron
microscope at 60 kV.

\subsection{Simulations} The computational simulations were produced
by custom software code written in the C programming language 
on the Linux operating system~\cite{mandrake}.

Random walk simulations were implemented by using the standard
pseudo-random number generator of the operating system. The
pseudo-random number generator was tested separately to ensure that
the degree of randomness of the generated data was adequate for our
purposes.

The simulations were compiled and run on the PIII Linux cluster of
the Department of Physics and Astronomy of The University of British
Columbia~\cite{p3cluster}.

\section{RESULTS}
\subsection{$\Ca$ oscillations}
In the rabbit IVC-ring preparation, laser scanning confocal microscopy was 
applied to image subcellular $\Ca$ waves of individual smooth muscle cells. 
Snap-shot images of a field of smooth muscle cells taken from the regions
enclosed by two white squares, from which a set of time-series images of
$\Ca$ fluctuations was derived, are illustrated in Fig.~\ref{kbr}A.

A representative trace of wave-like $\Caconin$ oscillations in response to 
phenylephrine (PE) is shown in Fig.~\ref{kbr}B. 
These oscillations were abolished by the application of 10 $\muM$ KB-R7943, a 
NCX reverse-mode  inhibitor~\cite{kbr}. We have previously shown that KB-R7943 
is highly selective for the reverse mode of NCX in the rabbit IVC~\cite{lee01}.
As shown in Fig.~\ref{kbr}, KB-R7943 causes the oscillations to cease only 
after a finite time delay, during which the amplitude of the oscillations 
declines, suggestive of depletion of $\Ca$ in the SR. 
In contrast, blocking the IP$_3$R-mediated release by administration of 2-APB, 
immediately abolished the $\Caconin$ oscillations~\cite{lee01}.
\bfig\centering
\includegraphics[scale=0.5]{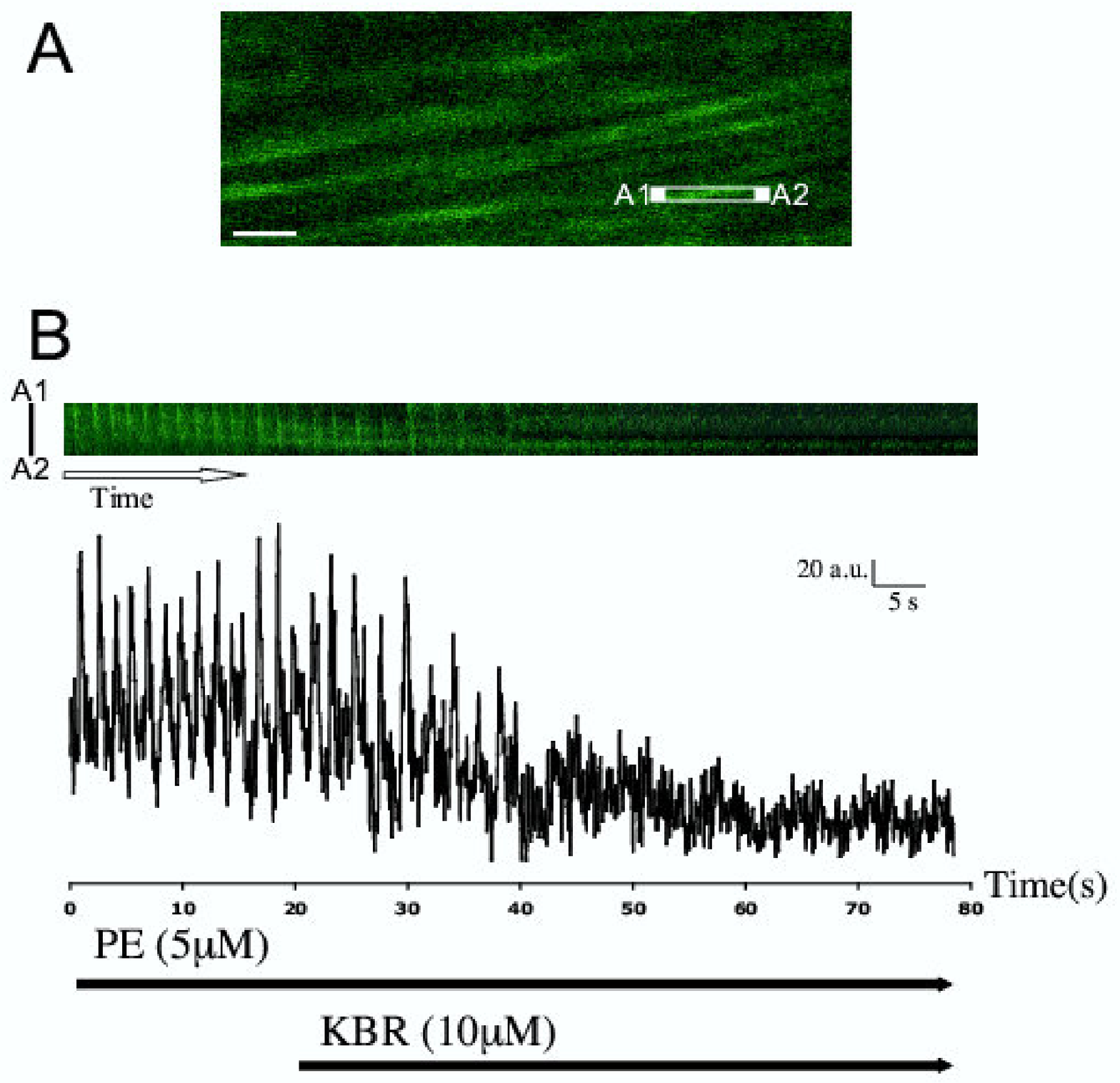}
\caption{
Phenylephrine (PE)-induced asynchronous wave-like $\Caconin$ oscillations in the rabbit IVC.
A: In intact rabbit IVC preparation, snap-shot images were acquired using
laser scanning confocal microscopy to image subcellular $\Ca$ waves of
individual smooth muscle cells. PE (5~$\muM$) elicited a rapid rise in
$\Caconin$  that appeared in the form of asynchronous $\Ca$ waves. $\Ca$
waves originated from distinct intracellular loci of fluorescence and
traveled along the longitudinal axis of long ribbon-shaped smooth muscle
cells. The region between the two white squares (A1 and A2) was selected to
record the fluctuation of $\Caconin$. The white scale bar in this panel
corresponds to $10\,\mum$. 
B: The fluctuation of $\Caconin$ between A1 and A2 was recorded over a time 
period of 80 s and plotted according to a time series. In the presence of PE, 
the region between A1 and A2 demonstrated fluctuations of $\Caconin$ to form 
wave-like oscillations over time. The PE-induced $\Ca$ waves were abolished in 
the presence of KB-R7943, an NCX reverse-mode inhibitor.}\label{kbr}
\endfig

\subsection{Ultrastructure}\label{microscopy}
By the methods described in Sec.~\ref{methods}, we have first generated a sequential series of images from 80-nm thick slices of a smooth muscle cell. A sample image from one of the sequences is reported in Fig.~\ref{stack}.
\bfig[t]\centering
\includegraphics[scale=0.4, angle=90]{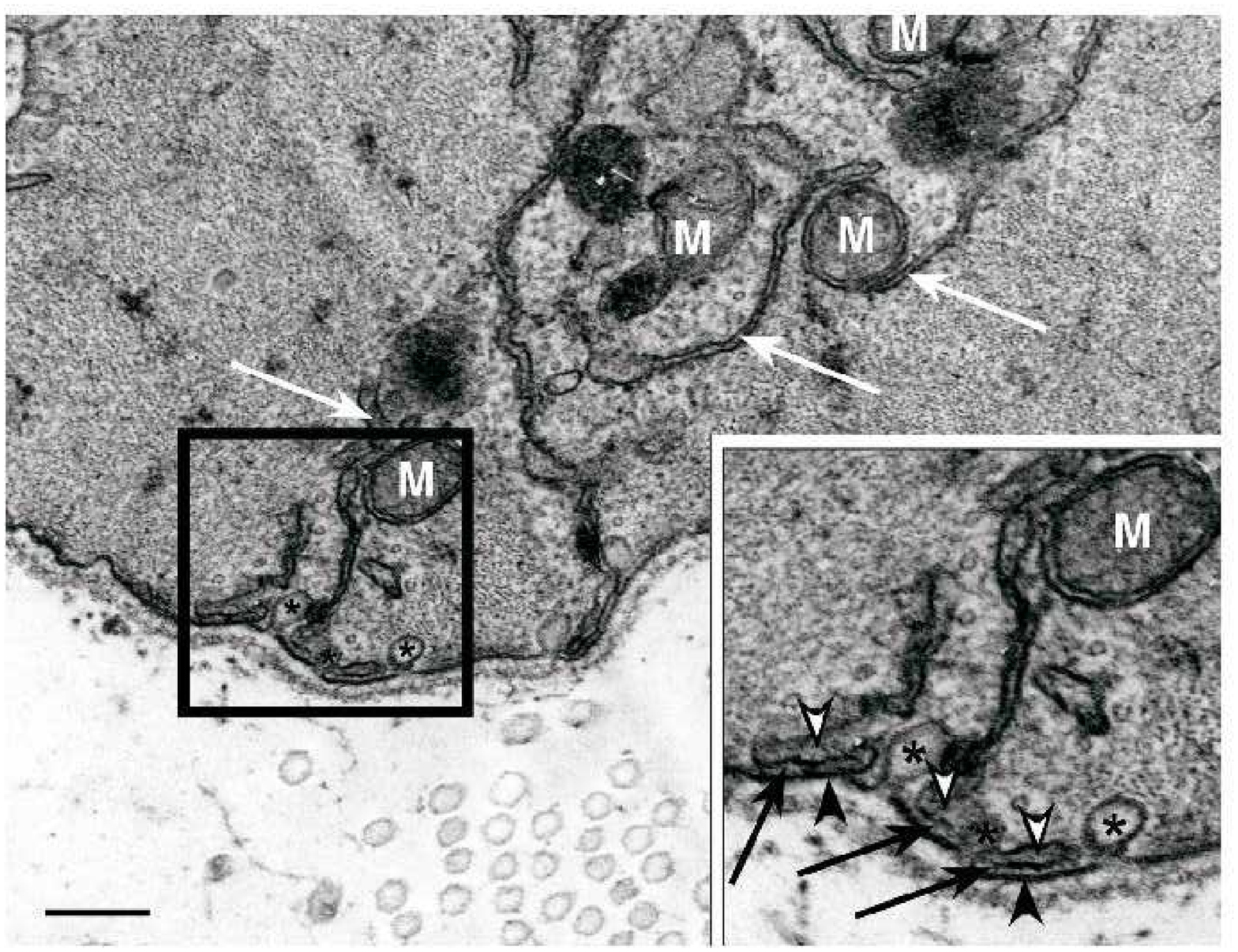}
\caption{
Representative image of serial sections
of \smc\ from
rabbit IVC obtained by transmission electron microscopy. 
This is one of the serial-section images used to generate three-dimensional
images of PM-SR junctional regions of vascular smooth muscle cells. The
inset on the lower-right corner is a magnified image of the area enclosed
by the black square. Superficial SR (indicated with white arrow heads) is 
observed to be apposed to PM (indicated with black arrow heads) and is flanked by two adjacent caveolae (indicated with black asterisks). In the core region of the vascular smooth muscle cell, deep SR (indicated with white arrows) can be found to be associated with mitochondria (indicated with M). In the inset, the PM-SR junction (indicated with black arrows) can be fully appreciated at high magnification. The black scale bar corresponds to 200 nm.}\label{stack}
\endfig
Sets of these images were then stacked to add depth information to the two-dimensional sections. With the use of high quality conventional electron micrographs (Fig.~\ref{stack}), we were able to observe the PM-SR junctional cytoplasmic space, located between the superficial SR and the apposing PM. The PM-SR junctions were often found to be associated with adjacent caveolae. The gap between the superficial SR and the apposing PM in PM-SR junctions measured approximately 20 nm. Regarding measurement of PM-SR junctions, in the present study 2 randomly selected PM-SR junctions per cell cross-section were measured from 30 different cells, corresponding to up to 60 junctions. With this unique spatial arrangement of caveolae flanking the 20 nm gap created by the superficial SR and apposing PM, the PM-SR junction forms a relatively restricted space that allows ions to accumulate to high concentrations. In order to demonstrate the three-dimensional structure of the PM-SR junction, we pseudo-colored the images and then combined the stacked serial sections and rendered them by applying a ray tracing algorithm (implemented via the Image J software package~\cite{imagej}) to generate pseudo-colored three-dimensional reconstructions of the PM-SR junctional regions of our sample cell. 
(Briefly, the ray tracing algorithm is a specific rendering algorithmic
approach of three-dimensional computer graphics included in the Volume J plug-in of Image J, where mathematically modeled visualizations of programmed scenes are produced using a technique which follows rays from the eye-point outward, rather than originating at the light sources. It produces results similar to ray casting and scanline rendering, but facilitates more advanced optical effects, such as accurate simulations of reflection and refraction, and is still efficient enough to be of practical use.)
In Fig.~\ref{pseudocol3D}, we present four snap shots of three-dimensional
images at different angles with respect to the z-axis for every 90 degrees. 
The PM is visible
in red, the SR in green, while white arrows indicate the PM-SR junction.
In panel A, the SR is situated in a position further away from the PM so that 
the plasmalemmal side of the superficial SR may be observed.
\bfig[tb]\centering 
\includegraphics[scale=.4]{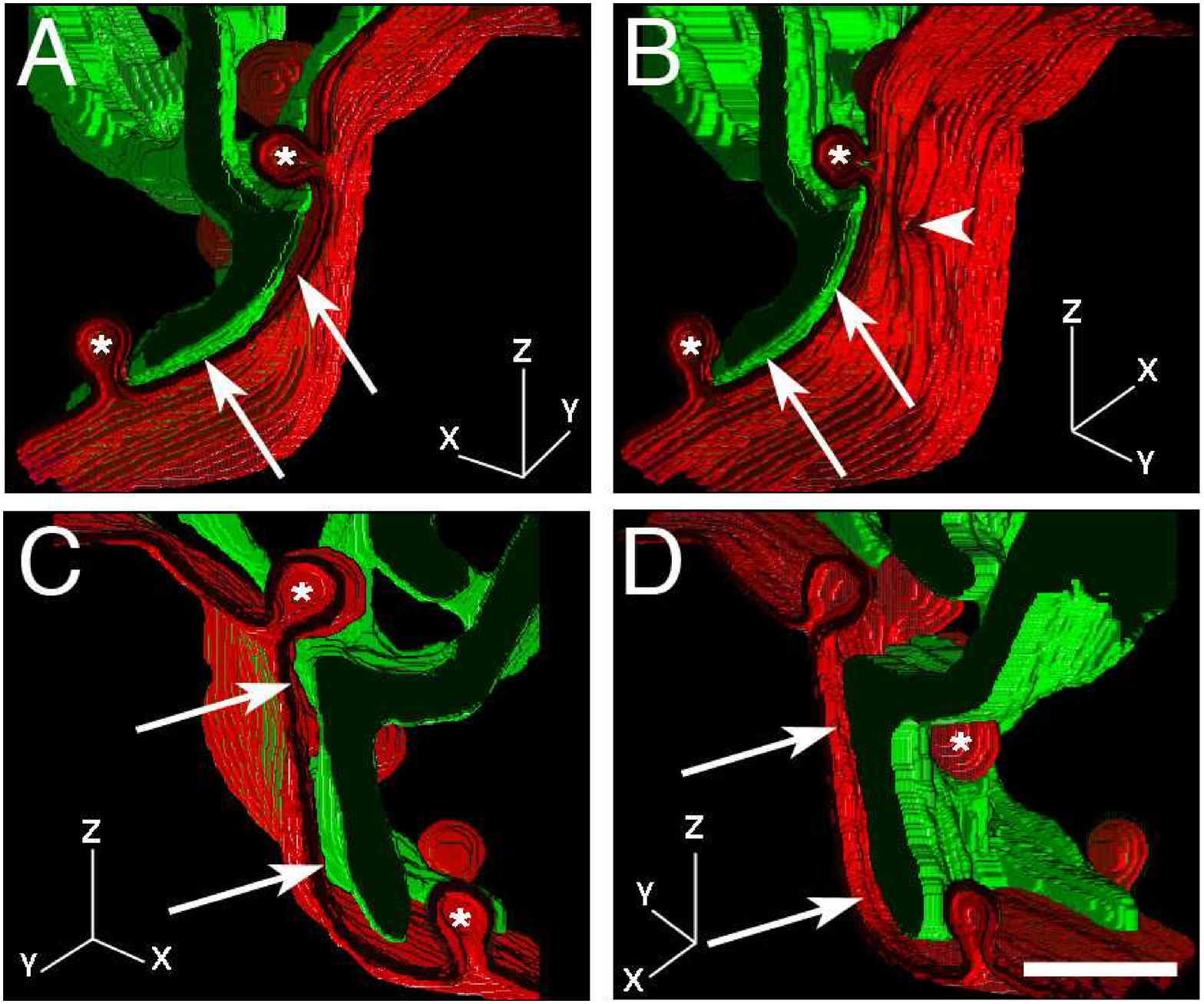}
\caption{
Pseudo-color recreation of PM-SR
diffusional barrier
(junctional space) regions of rabbit IVC \smc. The PM is
represented in red and the SR in green. The PM-SR
junctional space regions are indicated by white arrows, and caveolae
by white asterisks. The white arrowhead points to the mouth of a caveola in
the PM. The white scale bar in
panel D corresponds to 50~nm. The images in this
figure are screenshots taken from the animation provided as online
supplement to this article and they are four different views of the
three-dimensional 
reconstruction.}\label{pseudocol3D}
\endfig

In panel B, the view demonstrates the plasmalemmal side of the PM. The
arrowhead indicates the opening of a caveola on the PM. The view of panel
C is in a position that is rotated 180 degrees with respect to the 
z-axis of panel A. The image of panel C also provides a cross-sectional 
view of the
junctional space, as indicated by the arrows. Two caveolae, as indicated by
the asterisks, are associated with the SR and flank the lateral opening of
the junctional space. In panel D, the view is from the cytoplasmic side of
the SR. The PM is situated in a position further away from the SR allowing
for the cytoplasmic side of the PM and SR to be visualized as well as the
protrusion of the caveolae through the superficial SR into the cytoplasmic
space. 

To investigate localization of the NCX transporters, the
immuno-electron microscopy technique was applied to label the NCX
transporters. In an effort to preserve antigenicity and increase the
readability of the gold particle localization, muscle cells were
processed with a milder fixative, compared to the fixatives used for
the conventional electron microscopy study (as described in the
Methods section).  As can be noticed in Fig.~\ref{AuLabel}, the use of
a mild fixative inevitably compromised the preservation of cellular
morphology detail; however, sufficient detail was preserved to
determine the localization of gold particles. We measured the surface
density of gold particles in the PM to be  $(100\pm35)\,\mum^{-2}$ in
regions belonging to PM-SR
junctions, while it is $(4.0\pm1.3)\,\mum^{-2}$ in regions of the PM
not facing the SR.
\bfig[tb]\centering
\includegraphics[scale=0.4, angle=90]{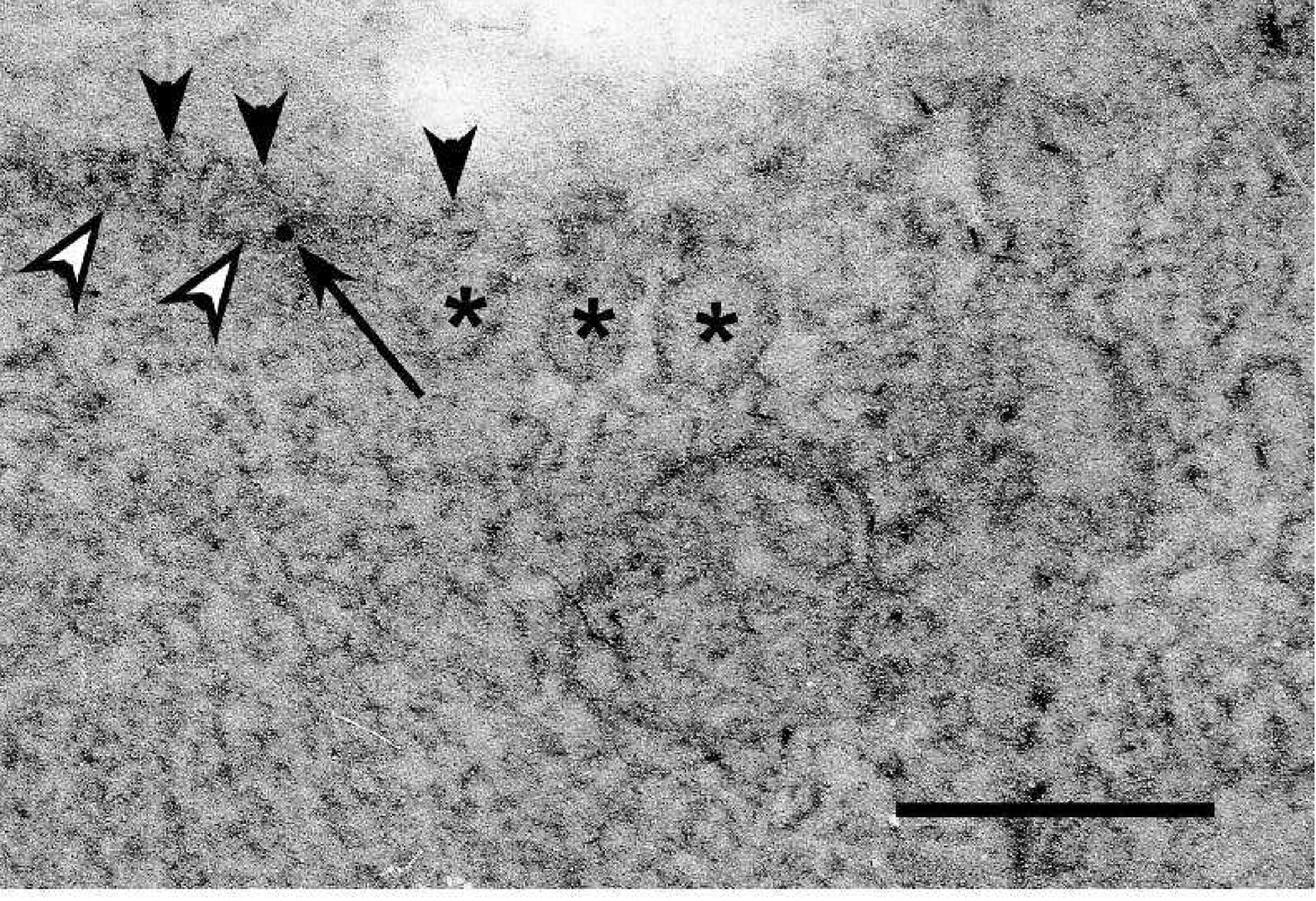}
\caption{
Immunogold labeling of NCXs: caveolae
are indicated by black asterisks, the SR is indicated by white
arrowheads and the PM by black arrowheads. The gold
particles (identified by black arrows) are recognizable by their
sharp edges. The black scale bar in this figure corresponds
to 100~nm. See text for explanation of the reasons for poorer
resolution of this kind of images compared to the high membrane
resolution ones displayed in Fig.~\ref{stack} }\label{AuLabel}
\endfig

\subsection{Model foundation}
Two main sets of information constitute the basis of our computational
model. Firstly, a survey of the literature indicates that for each
full contractile activation
of a typical smooth muscle cell, 
the amount of  $\Ca$  supplied to the myoplasm ($\Delta\Ca$) is of
the order of  $100\,\mumol$ per liter of cells~\cite{casey77,
ganitkevich96, ganitkevich00}. In our PM-SR junctional refilling
model,
this quantity is released from the SR while the SR is replenished via
$\Ca$ entry through the PM-SR junction~\cite{damon04}.  The second
information set relates to ultrastructure of the PM-SR junctions,
including their incidence, typical
geometrical shapes and dimensions~\cite{lee02-2}.
Electron microscopy revealed that about 15\% of the 
 \smc\ PM is closely apposed by the SR.
Furthermore, these images suggest that the typical junction is 
about 20~nm in height and extends in two dimensions for about 400~nm.

The idea then is that to replenish the \sr\ during the development of
a full contraction,
about
$100\,\mumol$ of $\Ca$ per liter of cells must enter the cell from
the extracellular space
(mainly through the NCX), traverse the junctional spaces, and
enter the SR via the SERCA pumps on the SR surface facing the PM. 
This number may be regarded as an upper limit as it assumes for
simplicity of
the model that there is no direct recycling of $\Ca$ between the
cytoplasm and the SR.

Using the typically observed and reported sizes and shapes, we model
the \smc\ as a long, thin cylinder about
$150\,\mum$ long and $3\,\mum$ in diameter (Fig.~\ref{SMCmodjctn}). 
\bfig\centering
\includegraphics[scale=0.3]{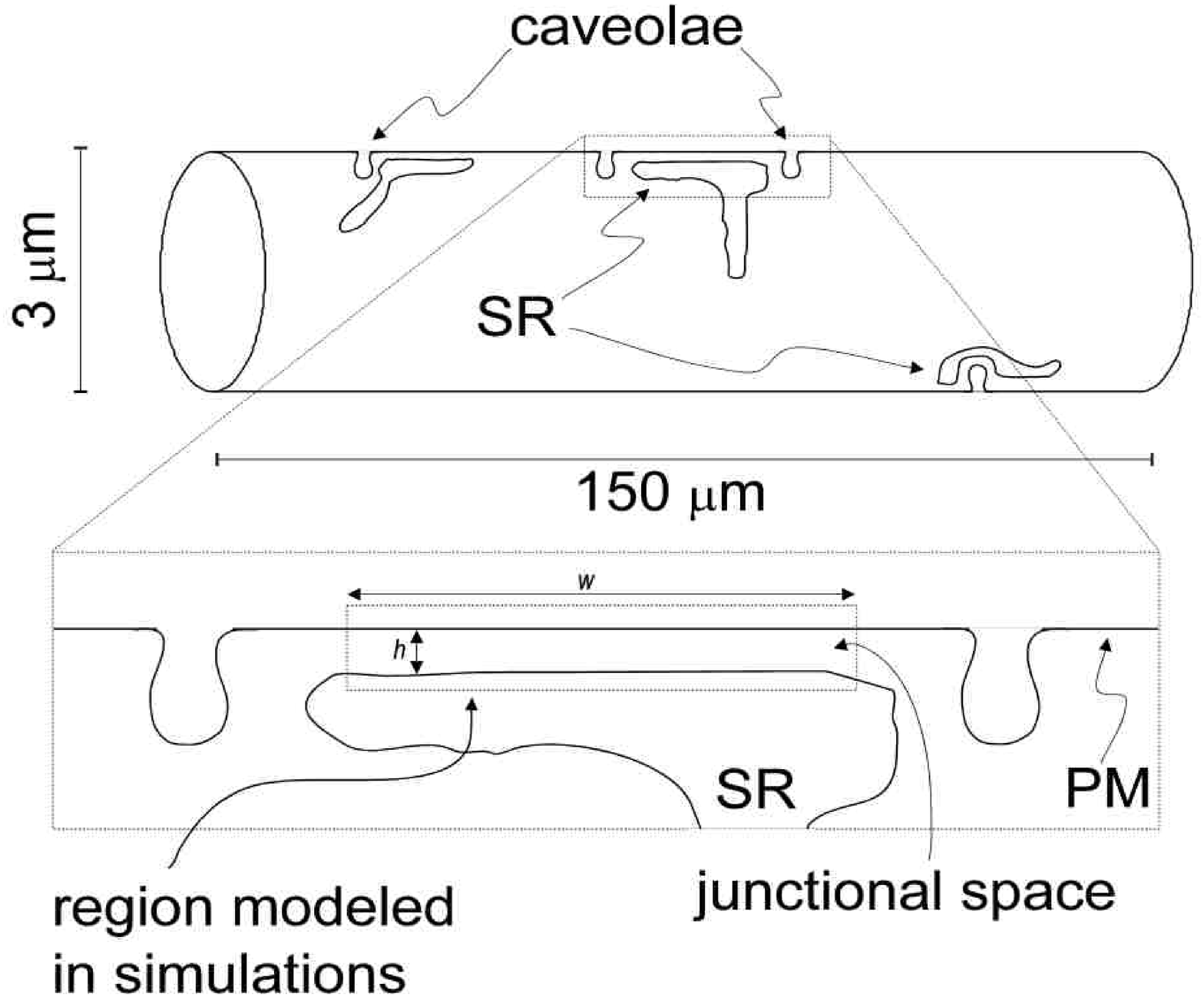}
\caption{
Diagram of smooth muscle cell and
diffusional barrier
regions considered in our model, where {\it h} and {\it w} are the height and width of the junction determined, on average, from ultrastructural images as 20 nm and 400 nm, respectively.}\label{SMCmodjctn}
\endfig
We furthermore estimate the average number of 
junctions in a cell, by recalling that: 1) the 
fraction of PM surface facing junctional SR 
($A_{\Sigma\rm JS}$) is 15\% of the \smc\ surface ($A_{\rm SMC}$),
which from 
the dimensions above is approximately $1.4\times 10^{-9}\,\m^{2}$,
hence
$A_{\Sigma\rm JS}=2.1\times 10^{-10}\,\oneun m^2$; 2) the surface
area of the portion of PM 
apposing {\it one} typical junction is approximately $A_{\rm 
JS}\simeq 1.6\times 10^{-13}\,\m^{2}$. Then, the ratio 
$A_{\Sigma\rm JS}/A_{\rm JS}$ produces an estimate of the average
number of 
junctions per \smc: $N_{\rm JS}\simeq 1300$ per cell.
The number of ions corresponding to the $\Delta\Ca$ of 
$100\,\mumol$ per liter of cells and which is 
expected to traverse those roughly 1300 junctions in each cell during
the 
occurrence of each full contraction is: $N_{\Ca}=(100\mumol$ per liter$)\times N_{\rm 
A}\times V_{\rm SMC}\simeq 6.4\times 10^{7}$, where $N_{\rm A}$ is
Avogadro's number and 
$V_{\rm SMC}$ is the volume of a \smc\ in liters (about $1\,{\rm pL}$
for 
our chosen cell geometry). Hence, during each contractile activation,
we expect about 
$N_{\Ca}/1300\simeq 5\times 10^{4}$ ions crossing each junction.

From this estimate coupled with $\Ca$ oscillation data, we can
also calculate the number of ions traversing a junction per unit time
during contraction. $\Ca$ wave data from studies of the rabbit IVC
show that oscillations occur with a frequency of about
0.5~Hz~\cite{lee01}, or a period $T$ of 2~s. Furthermore, contraction
force measurements conducted in parallel with $\Cacon$ wave
observations indicate that it takes around 125 oscillations to
achieve full contraction (note that we have used Fig. 2B, dotted trace, in reference~\cite{lee01}, where nifedipine is present to block voltage-gated $\Ca$ channels). Assuming that cellular $\Ca$
influx happens only during the
second half of each oscillation period, we then estimate that there
will be $5\times 10^{4}/(125\times T/2)\simeq 400\,\Ca/\oneun
s$ traversing each junction during contraction.
This assumption was made in order not to underestimate the flux density as
each oscillation appears to have a release phase and a refilling phase.
By taking a fixed value of the $\Ca$ flux during each half of an oscillation, we are virtually describing the $\Ca$ flux oscillations as a square wave, or a piece-wise constant, periodic function of time. The flux value during the second half of each $\Caconin$ oscillation is approximately an average value of the actual flux in the same time interval.

As anticipated in the introduction, earlier observations from this
laboratory established that this $\Ca$ flux is mainly mediated by
reverse NCX~\cite{lee01}. Thus, another important ingredient necessary
for the computational model is information on the typical number of
NCX per junction.
We described previously in section~\ref{microscopy} that
ultrastructural electron microscopic
images of \smc\ slices obtained with immunolabeled gold particles
yield a surface density of NCX in the junctional sections of the
PM of our cells. Transforming those measured densities to suit the
typical dimensions of the junctions observed (extending on average
400 nm in the two dimensions parallel to the PM), we can expect
$16\pm 8$~NCX per junction.

Another approach to the estimation of NCX density in our cells,
based on NCX turnover rate data, lends support to this finding.
Hilgemann has determined that {\it
maximal} NCX1.1 turnover rates ($V_{\rm max}$) in cardiac cells range from
1500~s${^{-1}}$ to 5000~s$^{-1}$~\cite{hilgemann96, hilgemann04}. It
is more difficult to find
reports of accurate measurements of this quantity for \smcs, where
the  NCX isoforms NCX1.3 and NCX1.7  are predominantly
expressed~\cite{nakasaki93, leesl94, quednau04}. 
Earlier, some
experimenters have reported NCX $V_{\rm max}$ values about one
order of magnitude
lower in bovine and porcine
arteries than the values known then for cardiac
cells~\cite{slaughter89, docherty95}.
For the purposes of this discussion, we will assume that the same
difference in
turnover rates between cardiac and \smc\ NCX applies to our system. 
Under the physiological conditions present during the \sr\
refilling phase observed in the rabbit IVC, it is plausible to expect
a $\Na$ concentration ([$\Na$]) around 20 mM, which is indeed
sufficient to switch the NCX to $\Ca$ influx mode, but just barely
(we discuss this matter further in section~\ref{discuss}). It is then
likely that the exchangers will be working at a much lower turnover
rate than maximal. From data on cardiac studies on the peak outward
current of the NCX1.1 as a function of [$\Na$]~\cite{hilgemann92}, we
estimated that the turnover rate at the conditions expected to
prevail in our junction is
only a few tens per second. It is then clear that 10 to 20 NCXs
per junction will suffice to provide the ionic rate
calculated above.

It is interesting to use the calculations above together with data on
dimensions of the typical \smc\ junctions to build a picture of the
crowdedness of our model \jn\ in terms of $\Ca$. 
Representing a typical junction with a parallelepiped of side
400~nm and height 20~nm (hence a volume of
$3.2\times10^{-18}$~L)~(Fig.~\ref{ModJnctn}), the previously
calculated flux per cycle yields a junctional $\Ca$ concentration of
about $200\,\muM$, greatly exceeding the bulk $\Caconin$ (the latter is typically around 100~nm~\cite{alberts02}). Under these
conditions, assuming a $\Ca$ is a
little
sphere of radius equal to the atomic Bohr radius of calcium (0.2~nm),
 then,  if each
ion were the size of an apple, the next nearest ion would on average
be more than three meters away. This $\Ca$ density in the
junctions is sufficiently low that we can attempt to
describe their trajectories by treating ions as
individual units, moving through the \jn\ in a random walk fashion
due to diffusion in the junctional cytosol.

The data used to obtain the quantities estimated in this section are
summarized in Table~\ref{data}.

\begin{table}
\begin{center}
    \begin{tabular}{rlc} \hline\hline
    & & References\\\hline
	SMC diameter & $3\,\mum$ & \cite{somlyo75}\\
	SMC length & $150\,\mum$ & \cite{somlyo75}\\
	jPM-jSR separation & ($20\pm 0.4)\,\nm$ & this article\\
	junction lateral extension & ($382\pm 67)\,\nm$ & this article and
\cite{lee02-2}\\
	surface area ratio jPM/PM & $15\%$ & \cite{lee02-2}\\
	$\Ca$ release/full contraction & $100\,\mumol$ per liter of cell
	& \cite{casey77, ganitkevich96, ganitkevich00} \\
 estimated $\Ca$ influx for contraction & 400 $\Ca$/s/junction & this article \\
	frequency of $\Caconin$ oscillations & $(0.5\pm 0.02)$ Hz &
\cite{lee01}\\
	time to maximum force & 250 s & \cite{lee01}\\
	Number NCX/junction & $16\pm 8$ &  this article\\ 
	\hline\hline
    \end{tabular}
    \caption{Summary of data used in the model.}\label{data}
\end{center}    
\end{table}

\subsection{Computational Model}\label{compmodel}
As indicated earlier, electron microscopy of immunogold labeled
proteins yielded
information on the density of NCX in \smc\ \jns.
On the other hand, information on the density of SERCA pumps on the
\sr\ surface side of \jss\ is not as easily available, nor as
accurate. The maximum number of pumps to distribute on the jSR
of the model junction is determined using estimations of the
number of pumps per cell from data in the literature and approximate
measurements of the
ratio between jSR and total SR obtained from our electron micrographs
of IVC \sm\ cells. 

With the above information, the geometry
of our numerical simulations can be set up as illustrated in
Fig.~\ref{ModJnctn}, where the parallelepipedal shape of the model
junction is arbitrarily chosen in the interest of ease of
implementation in the simulations (the exact chosen shape is a box with height of 20 nm and square bases with sides of 400 nm).
\bfig[t]\centering
\includegraphics[scale=0.3]{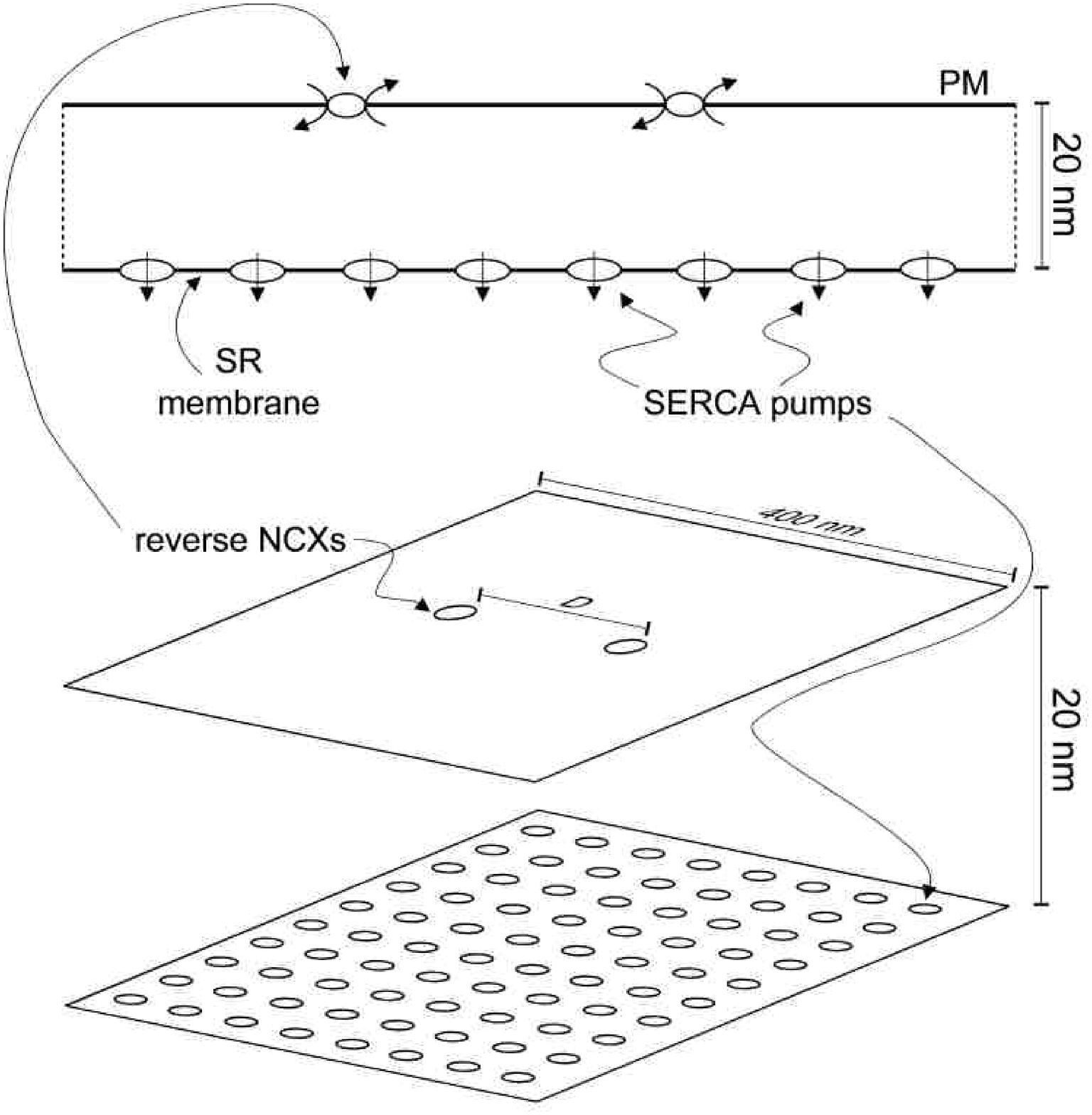}
\caption{
The model used to represent {\it one}
junction (not to
scale). The 20-nm separation between membranes and their 400-nm
lateral extension are estimated from electron micrographs. The
parallelepipedal shape is chosen for ease of implementation in the
simulations. In them, the model junction has 2 sources on the jPM, at a distance $D$ from each other. $D$ varies from 0 (i.e., only one
source at the center of the junctional PM (jPM)) to 400~nm (i.e., two
sources at opposite edges of the jPM), while the number of sinks is left as a variable (see text for details).}\label{ModJnctn}
\endfig
While on the one hand we know with a good degree of reliability that in the observed oscillations $\Ca$ is transported (principally) from NCX on the jPM to SERCA on the jSR, on the other, the process taking the ions from sources to sinks is very complex. It includes thermal agitation, collisions with other molecules and ionic species of the junctional cytosol, interaction with jPM and jSR and buffering/unbuffering of $\Ca$ by junctional protein structures.\footnote{The issue of junctional $\Ca$ buffers is by no means a trivial one and it has been the subject of extensive theoretical studies (see for example \cite{sneyd94, compcellbio}).To a first level of approximation, the general effect of $\Ca$ buffers can be described by the introduction of an ``effective'' diffusion coefficient into the ion transport equations.}
Ionic trajectories under these conditions are
virtually impossible to resolve deterministically by solution of the
equations of motion of the individual ions.
Instead, in this computational model for
$\Ca$ transport we propose that ions move around and diffuse in the
junctional cytosol in a process akin
to that of a gas expanding to occupy a given available volume. The term ``diffuse'' is used here rather loosely to encompass the set of actions causing $\Ca$ entering the junctions via the NCX to make their way to the SERCA pumps.
It is then possible to represent the ionic trajectories as a
random walk in three dimensions~\cite{berg93, whitney90}. 
It should be emphasized that based on previous
findings by this laboratory, we can exclude in this first version of
the model that the activity of other transporters is of much
importance to SR refilling during maintenance
 of the asynchronous $\Ca$ waves studied here.

These trajectories are simulated by randomizing ionic positions 
starting from their entrance into the \js\ through point sources,
(NCX in Fig.~\ref{ModJnctn}), to 
either their capture by a sink (SERCA in Fig.~\ref{ModJnctn})
or their exit from the \js\ through its sides, which are open to the bulk cytosol.
The random walk is executed on a cubic lattice of 0.2-nm steps, a value 
dictated by the need to compromise factors like the estimated mean free path 
of $\Ca$ in the junctional cytosol, the average path length of an ion from 
source to sink in the junction, the values for the diffusion coefficient found 
in the literature (from about 10 to about 200 $\mum^2/{\rm s}$ as reported 
in \cite{allbritton92}), and the need to achieve a realistic enough 
description of the 
random walk, while still keeping the simulation time reasonable. 
The typical distance travelled by a random walking ion and the available 
diffusion 
coefficient values yield an
average travel time ranging from about 10 to about 100 $\mus$.
We have used a fixed time step value of the order of 10 ps
determined from the total expected junction traversal time for a typical
$\Ca$ simulated trajectory and an average number of random walk steps taken
for the traversal (the value of the diffusion coefficient $D$ obtained as
$D=({\rm space\; step})^2/(6\times{\rm time\; step})$ is indeed of the same 
order of 
magnitude of the
values found in the literature).
The position 
of an ion, say the {\it n}-th ion, is defined by an $(x_{n}, y_{n}, z_{n})$ 
triad beginning from an initial position $(x_{n}^0, y_{n}^0, z_{n}^0)$ 
coinciding with an ionic source (NCX): $(x_{n}^0, y_{n}^0, z_{n}^0)=(x_{s}, 
y_{s}, z_{s})$ (where ``s'' stands for ``source''). At each successive step ions can move with equal probability in any of the 6 directions allowed by the lattice and the new ionic position is chosen by changing the $x_n$, $y_n$, and $z_n$ values each by 0.2 nm either positively or negatively in a random fashion, until the boundary conditions are met.
In the simulations, ions encountering the jSR
away from the SERCA or the jPM simply bounce of these surfaces and
continue on with their random walk (perfect reflecting boundary conditions at jPM and at jSR away from the sinks). Also, ions reaching the lateral
boundary of the junction are considered lost as far as the capture
count is concerned (perfect absorbing conditions at lateral boundary).
Sink radii are estimated at 4.5~nm from structural information on
SERCA pumps obtained from the ``Protein data bank''~\cite{pdb,
pdbsite, sercastructure}. 

We can then infer the probability that ions are captured by the
sinks, by taking the ratio of the 
number of captured $\Ca$ over the total number of ions having entered
the junction during one full contraction. The variable called ``captures''
in the graphs reported in Fig.~\ref{varncxD}, \ref{ncxsnks} and \ref{varht} is
calculated in this way. The simulations are meant to represent a snap-shot of the captures at the end of 1 s of oscillation, i.e., the half of the oscillation corresponding to the refilling of the SR. From the $\Ca$ influx in that phase of the oscillations estimated in the previous section, 400 $\Ca$ random walk through the junction and the capture probability is calculated. To obtain a ballpark figure for the uncertainty of these calculations, each datum in the graphs is the mean value from 10 repetitions of the simulations and the error bars shown correspond to 3 standard deviations from the mean.

In the proposed model, while we have an estimate of the number of
sources in the junctions, we have no firm information on their
position in the jPM. Moreover, there is currently very limited
knowledge about the shape of the $\Naconin$ gradient inside the
junctional spaces. It is therefore very difficult to estimate how
many of the total estimated number of junctional NCX will be active in
$\Ca$-influx mode during each contraction-causing $\Caconin$
oscillation. Because of this presently unresolved uncertainty, we
have chosen to implement only two sources in our model (i. e., to
consider that only two of the junctional NCX reverse) to explore two
of its aspects: 1) the plausibility of the model ideas, as it is
reasonable to think that if the model works with only two sources, it
will work {\it a fortiori} with a larger number of sources; 2) the
dependence of the capture rate on the relative position of the
sources: the anticipated uncertainty in the details of the junctional
$\Naconin$
gradient and the possible proximity of some of the NCX to
the edge of the junction lead us to expect that the positioning of
the NCX on the jPM bear influence on the capture rate. In our
computation, we have first
studied the latter item by placing the sources at several distances
from each other ($D$ in Fig.~\ref{ModJnctn}) along an axis
bisecting the jPM  and calculating the
capture
probability under these configurations. The results are reported in
Fig.~\ref{varncxD}. For this simulation we placed 100 sinks on the
jSR in a regular square grid pattern (this is the way SERCA on the jSR are organized in all simulations presented here). This number is an educated guess arrived at with the benefit of
hindsight via information on the turnover rates and on the reported
density of
SERCA pumps in SR membrane, as is explained in more detail in the
next section.
\bfig[tb] \centering
\includegraphics[scale=0.5]{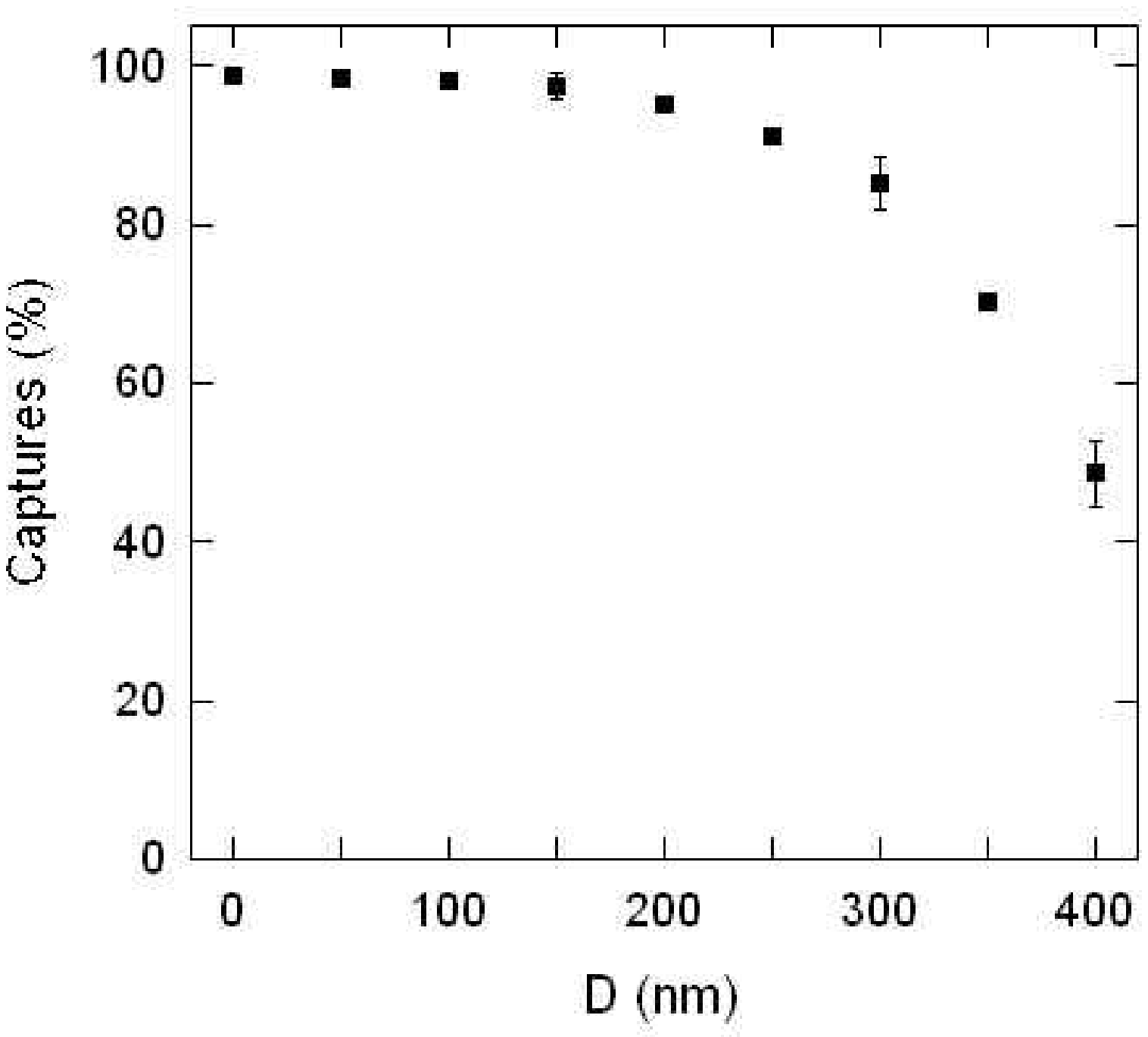}
\caption{
Results of simulations based on the
model junction of Fig.~\ref{ModJnctn}. The ``captures'' are the
percentage of $\Ca$
landing on a sink (SERCA) after entering the junction from one of two
sources (NCXs) and random-walking through the junction. $D$ is the
inter-source distance; a fixed number of 100 sinks was placed in the jSR
for these simulations. A few representative error bars are shown. They
correspond to 3 standard deviations from the mean value of 10 repetitions
of the simulations at each value of $D$. (See online supplemental files for
virtual three-dimensional view of simulation setup.)}\label{varncxD}
\endfig
However, given the uncertainty in the number of pumps on the jSR, we
have also calculated the capture probability {\it vs.} the number of
sinks for our model junction. The results are shown in
Fig.~\ref{ncxsnks}, where each set of data (indicated by different symbols
and colors as described in the legend) represents the result of simulations
with a different value of the inter-source distance $D$. The files supplied
as online supplement to this article depict to-scale virtual
three-dimensional views of the simulation setup, along with two typical traces of $\Ca$ paths.
\bfig[tb] \centering
\includegraphics[scale=0.5]{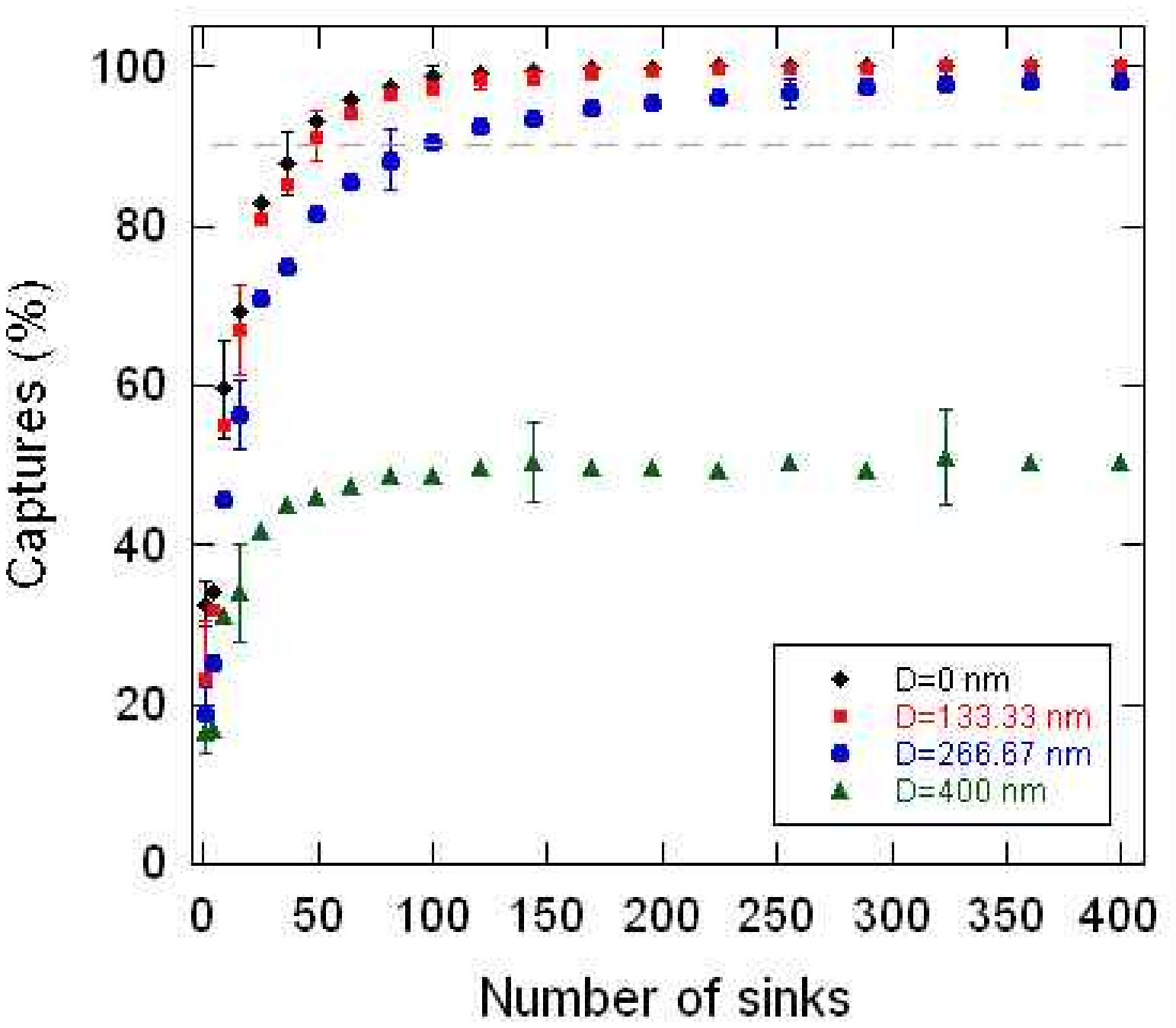}
\caption{
Results of simulations based on the model junction of Fig. \ref{ModJnctn}.
The ``captures'' are the percentage of $\Ca$ landing on a sink (SERCA)
after entering the junction from one of the two sources (NCX) and
random-walking through the junction. The dashed line indicates the 90\%
capture rate value, which is achieved on average with about 100 sinks
(except for $D=$400 nm). Each datum in this plot is an average of 10
simulations and each set of data, indicated by the different symbols in the
legend, is obtained with a different value of the inter-source distance $D$
(Fig. \ref{ModJnctn}). Error bars represent 3 standard deviations from the
mean of 10 simulations.(See online supplemental files for virtual
three-dimensional view of simulation setup.)}\label{ncxsnks} 
\endfig

A striking property of the PM-SR junctions is the consistency of the
20-nm width of the junctional cytoplasmic space. When this width has
widened, as seen during treatment with calyculin A, the $\Ca$ waves
ceased to exist indicating that it is a critical feature for linked
$\Ca$ transport~\cite{lee05}. To examine this hypothesis we
 studied the effect on the capture rate of varying the PM-SR
separation of our model junction.
For this we have carried out another type of simulation, placing 2
sources in the jPM, 100 sinks in the jSR and computed the capture
rate (as defined before) at different values of the jPM-jSR separation,
from 0 to 400 nm. Fig.~\ref{varht} illustrates the outcome of this
computation, in which a random walk lattice step of 0.1 nm was used for the capture data at jPM-jSR separation values from 1 to 50 nm.
\bfig[tb] \centering
\includegraphics[scale=0.45]{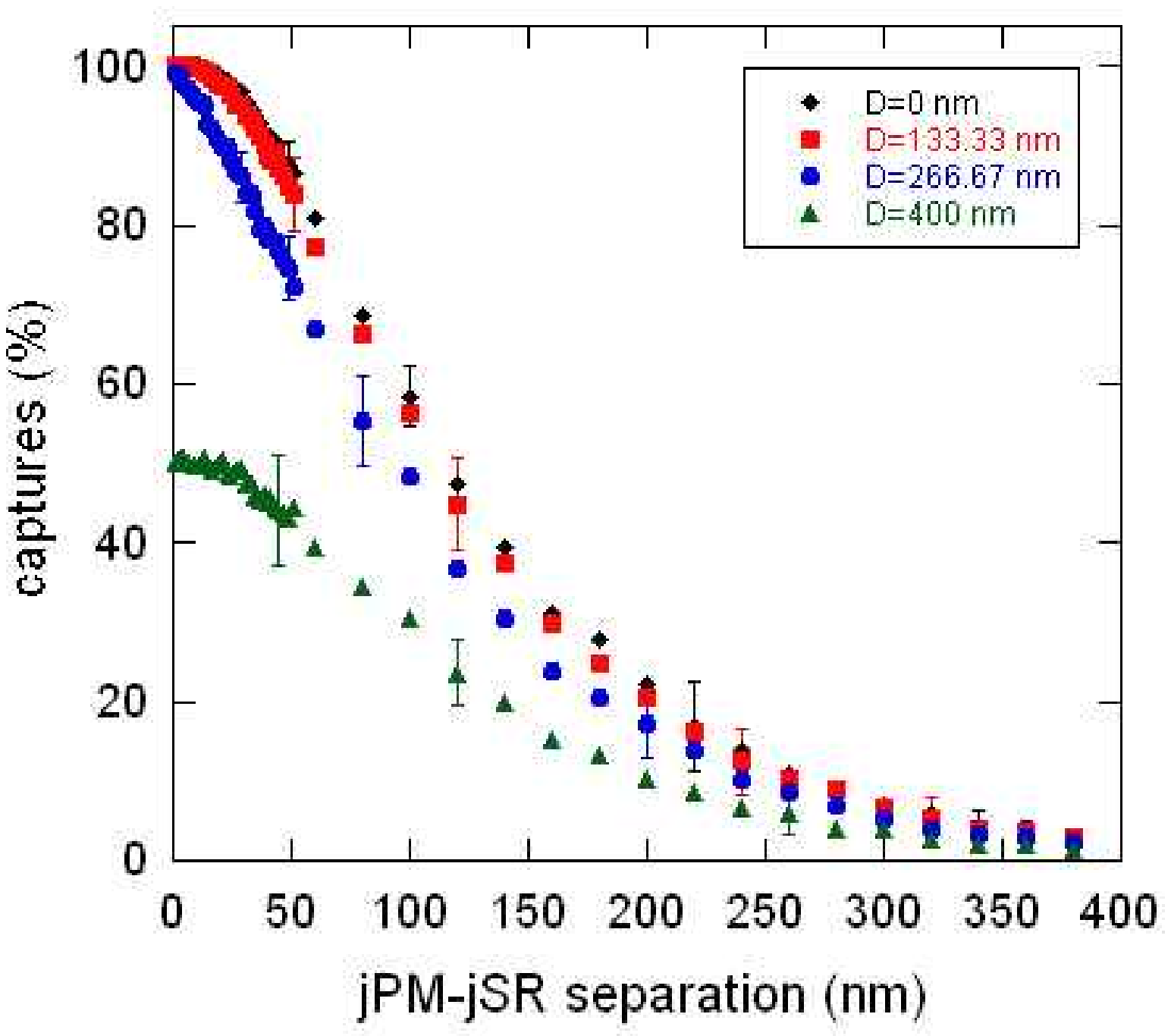}
\caption{
Results of simulations based on the
model junction of Fig.~\ref{ModJnctn}. The capture rate is calculated
at different values of the distance between the jPM and the jSR, with 100 sinks in the jSR. 
As in the previous graph, each datum in this plot is an average
of 10 simulations and each set of data refers to a different value of the inter-source distance $D$ (Fig. \ref{ModJnctn}). The capture data for PM-SR distance values between 1 and 50 nm were obtained with a random walk lattice size of 0.1 nm.}\label{varht}
\endfig

\section{DISCUSSION}\label{discuss}
This work combines physiological studies on asynchronous $\Cacon$ 
oscillations observed by confocal microscopy in
phenylephrine-stimulated rabbit IVC and 
electron microscopy studies of the subcellular ultrastructure of the
same cells to 
build a foundation for a computational model that describes the
refilling 
process of SR with $\Ca$ to maintain the excitatory $\Ca$ waves. As
such it also provides a novel computational approach for
simulating the cytoplasmic $\Ca$ gradients, which
are the basis for site-specific $\Ca$ signaling in the vasculature.

The specific question we set out to address by the development of this
model was whether it would be possible to move sufficient $\Ca$ to
sustain contraction through the PM-SR junctions alone, occupying
only about 15\% of the cell membrane surface area. 
Since the bulk of the refilling process is mediated by linkage of
$\Ca$ entry through reverse NCX to SERCA, we decided to model this
aspect of the oscillatory $\Ca$ cycle~\cite{lee01}. As such this
novel approach lays the groundwork for future incorporation of other
localized processes, such as NSCC mediated $\Na$ influx and $\Na$
extrusion by $\Na$/K$^+$ pump NKA$\alpha_2$. The
results of the numerical simulations based on the physical and
physiological features of our model cells suggest that the proposed
PM-SR junctional $\Ca$ transport is indeed realistic. In spite of the
scarce availability of
accurate experimental data on the SERCA 2b, this pump's
isoform preferentially expressed in  vascular \smc, and on the NCX
isoforms NCX1.3 and NCX1.7 (mainly, the maximal rates of both
transporters and the
density of the former are in short supply), the main 
simulation results
reported in Fig.~\ref{ncxsnks} are a strong indication that the PM-SR
junctional complexes present in the \smcs\ of the rabbit IVC are
capable of
achieving the capture rates of $\Ca$ by the SR necessary to sustain
the repetitive $\Ca$ release during excitatory $\Caconin$
oscillations.

The results of Fig.~\ref{varht} further highlight the possible need for the
principal transporters responsible for sustaining the observed
asynchronous $\Ca$ waves to be localized in close proximity to one
another by means of close apposition of the PM and the SR in
junctions. The simulation data show a curvature change at a
separation of 40 to 50~nm: below that value the curve appears concave downward, while it is concave upward elsewhere. This feature may be interpreted as follows. For junctional gap values below 40 to 50 nm, spillage of $\Ca$ to the
bulk cytosol is sufficiently small to guarantee efficient refilling
of the SR. If instead the PM-SR separation is greater than about
40 to 50 nm, loss of $\Ca$ from the junctional space to the bulk cytosol
increases rapidly and may explain the cessation of $\Caconin$
oscillations upon experimental separation of the PM from superficial
SR~\cite{lee05}. 

Efficient refilling also clearly depends on the presence of an
adequate number of target SERCA on the SR side of the junction.
Estimates of the density of SERCA pumps in smooth muscle cells from
data in the literature vary from 6 to 18 million pumps per
cell~\cite{inesi90, elmoselhi95, holmes00}. Assuming random
distribution, we can thus make a
ballpark calculation of the expected number of pumps per junction
by using 12 million SERCA per cell. Measurements from the
ultrastructural images we present herein suggest a ratio of about 1
to 30 between jSR and total SR. Having calculated that, on average,
there are 1300 junctions per cell, we arrive at an estimate of about
300 SERCA/PM-SR junction.
Simulation results (Fig.~\ref{ncxsnks}) from our model support this
figure, showing that between 200 and 400 SERCA pumps are necessary to
capture more than 95\% of the amount of $\Ca$ required to refill the
SR to maintain optimal smooth muscle contraction.

With respect to the NCX, its ability to reverse hinges on competition
between the junctional $\Caconin$ and $\Naconin$.
For the model to work the NCX has to reverse only at the PM-SR
junctions. This requires that only there the following relation be
obeyed between the electrochemical potential of $\Na$ ($E_{\rm Na}$),
of $\Ca$ ($E_{\rm Ca}$) and the membrane potential ($V_{\rm M}$)~\cite{blaustein99}:
\beq
3E_{\rm Na}-2E_{\rm Ca}\le V_{\rm M}\label{ncxrev}
\endeq

By expanding Eq.~(\ref{ncxrev}), recalling that in general $E_{\rm
C^{z+}}=RT/({\rm z}F)\,{\rm ln}([C^{\rm z+}]_{\rm out}/[C^{\rm
z+}]_{\rm in})$, where $R$ is the universal gas constant, $T$ the
absolute temperature, ${\rm C}^{\rm z+}$ is a z-valent cation, and
$F$ is Faraday's constant, we obtain a lower limit on the value of
$\Na$ concentration $\Naconin$ necessary to achieve NCX reversal. We
find such limit as $\Naconin > 12$~mM, using $V_{\rm M}=-60$~mV~\cite{Vm},
$\Naconout=145$~mM, $\Caconin=10^{-4}$~mM, $\Caconout=1.5$~mM (ionic concentrations from~\cite{alberts02}),
$R=8.3$~J/(mol K), $T=310$~K, and $F=9.65\times 10^4$~J/(V mol). 
At the same time it is understood that increases
in junctional $\Cacon$ will increase  the junctions $\Nacon$ for reversal, while depolarization
decreases it. Another
factor favoring a higher $\Nacon$ in the \js\ is that the low
affinity isoforms NKA$\alpha_2$ and NKA$\alpha_3$, having
$k_{\rm d}\approx$24--33~mM ($k_{\rm d}$ is the concentration of ligands at which 50\% of receptors are occupied), appear to be localized at the
junctions~\cite{blaustein99, juhaszova97}. The probability of NCX
reversal at the junctions is further supported by
measurements by our laboratory (unpublished data),
which suggest a sub-plasmalemmal $\Naconin$ approaching 20~mM under
stimulating conditions. Using data from patch-clamp  studies on $\Ca$
influx current of cardiac sodium-calcium exchange as a function of
[$\Na$], we estimated that at values of $\Naconin$ around
20~mM the NCX in our cell junctions would be transporting in the
reverse mode at about 30\% of
$V_{\rm max}$, or likely around 50~s$^{-1}$.

Clearly, it would lend considerable support to the model if the NSCC
and NCX were clustered at the PM-SR junctions and if $\Na$ removal
was regulated not to decrease $\Naconin$ below NCX reversal values.
By electron microscopy of gold immunolabeled smooth
muscle cells (Fig.~\ref{stack} and~\ref{AuLabel}), we have determined
the density of NCX transporters in junctional and non-junctional
regions of the PM. Our results indicate a twenty five-fold higher
density of NCX in the jPM compared to the non-junctional PM. 
Although there is strong evidence that NSCC play a major role in the
maintenance of asynchronous oscillations (\cite{lee01, lee02-1})
their possible localization at the PM-SR junctions remains to date an
intriguing hypothesis. Clustering of the NSCC would give rise to
non-uniform  gradients that would work as position-dependent
activators of $\Ca$-influx mode NCX. 

The present quantitative model for refilling of smooth muscle SR
during asynchronous $\Caconin$ oscillations at the PM-SR junctions
answers one important question and sets a solid stage for exploring
structure and function of junctional complexes between organellar and
cell membranes. The computational data presented in figures
\ref{varncxD}, \ref{ncxsnks} and \ref{varht} strongly suggest that it is
possible to refill the SR through PM-SR junctional complexes, which
occupy 15\% of the smooth muscle cell membrane and thereby support
our qualitative model derived from experimental data~\cite{lee02-2}.
Some of the urgent questions related to the proposed model of PM-SR
junctional $\Ca$ transport are: 1) What are the detailed
characteristics of the NCX  vascular \smc\ isoforms NCX1.3 and
NCX1.7, specifically in regards to their $V_{\rm max}$? 2) What is the
identity of the NSCC (to date it has been suggested that TRPC1 or
TRPC6 are involved)? 3) Where are the NSCC localized on the PM? 4) 
How do the positions of NCSS, NCX, SERCA and NKA shape the junctional
$\Na$ and $\Ca$ gradients to determine, together with the membrane
potential, which NCX reverse? 5) What are the turnover numbers and
localization of smooth muscle SERCA isoforms? 6) What is the nature
of the $\Ca$ binding  sites in the junctional space and their effect
on the diffusional paths of the ions? 7) Which molecules provide the
glue and spacing of the cytoplasmic junctional space? 
In more general terms the above quantitative approach to simulating
cytoplasmic $\Ca$ domains could help resolve important unanswered
questions related to
differential, time and space specific $\Ca$ signaling of myofilament
activation, mitochondrial metabolism, apoptosis, migration and
differentiation.
\vspace{10mm}

{\bf Acknowledgments}
\vspace{5mm}

We are indebted to Amir Ali Ahmadi for his help in the early phases
of this effort. We wish to thank Andre
Marziali and Jon Nakane 
(Applied Biophysics Laboratory, Department of Physics and Astronomy,
University of British Columbia) 
for their invaluable input into the modeling phase of the project, as
well as Damon Poburko for helpful discussions. The research was
supported by grants from the
Canadian Institute of Health Research and the Heart and Stroke
Foundation of British Columbia and Yukon.
N.~F. is supported by funds from the Child and Family Research
Institute (CFRI) of British Columbia. The simulations described here
were performed on the ``vn''
cluster of the University of British Columbia (supported, in turn, by
grants by the Canadian Foundation for Innovation, the British
Columbia Knowledge Development Fund and the Canadian Institute for
Advanced
Research). We would finally like to thank the James Hogg iCapture 
Center for Cardiovascular and Pulmonary Research (University of 
British Columbia) for their support.

\end{document}